\documentclass[dvips]{plain}
\usepackage{supertabular,lscape,epsfig}
\usepackage{amssymb}
\usepackage{amsmath}
\usepackage{wasysym}

\SetPages{1}{0}

\SetVol{0}{2020}

\begin{document}

\begin{Titlepage}
\Title{A Newtonian model for the WASP-148 exoplanetary system enhanced with TESS and ground-based photometric observations}
\Author{G.~Maciejewski$^{1}$,  M.~Fern\'andez$^2$,  A.~Sota$^2$,  A.~J.~Garc\'ia Segura$^2$}
{$^{1}$Institute of Astronomy, Faculty of Physics, Astronomy and Informatics,
         Nicolaus Copernicus University, Grudziadzka 5, 87-100 Toru\'n, Poland,
         e-mail: gmac@umk.pl\\
  $^2$Instituto de Astrof\'isica de Andaluc\'ia (IAA-CSIC), Glorieta de la Astronom\'ia 3, 18008 Granada, Spain}

\Received{December 2020}
\end{Titlepage}

\Abstract{The WASP-148 planetary system has a rare architecture with a transiting Saturn-mass planet on a tight orbit which is accompanied by a slightly more massive planet on a nearby outer orbit. Using new space-born photometry and ground-based follow-up transit observations and data available in literature, we performed modeling that accounts for gravitational interactions between both planets. Thanks to the new transit timing data for planet b, uncertainties of orbital periods and eccentricities for both planets were reduced relative to previously published values by a factor of 3--4. Variation in transit timing has an amplitude of about 20 minutes and can be easily followed-up with a 1-m class telescopes from the ground. An approximated transit ephemeris, which accounts for gravitational interactions with an accuracy up to 5 minutes, is provided. No signature of transits was found for planet c down to the Neptune-size regime. No other transiting companions were found down to a size of about 2.4 Earth radii for interior orbits. We notice, however, that the regime of terrestrial-size planets still remains unexplored in that system.}{planetary systems -- stars: individual: WASP-148 -- planets and satellites: individual: WASP-148 b, WASP-148 c}


\section{Introduction}

The WASP-148 planetary system comprises two Saturn-mass planets orbiting the Sun-like star TYC 3083-295-1 (H\'ebrard \etal 2020). The inner planet, WASP-148~b, was detected with the transit technique, producing flux drops of a depth of $7$ ppth (parts per thousand) every 8.8 days. The planet's orbit was found to be orientated almost perpendicular to the sky plane with an inclination $i_{\rm{b}} = 89.90^{\circ} \pm 0.27^{\circ}$. The orbital eccentricity was found to be non-zero at a $3.5\sigma$ level. The planet has a mass and radius equal to $0.291 \pm 0.025$ $M_{\rm{Jup}}$ (Jupiter masses) and $0.722 \pm 0.055$ $R_{\rm{Jup}}$ (Jupiter radii), respectively. Its orbital motion is perturbed by the outer planet WASP-148~c that resides on a 34.5-day eccentric orbit. It was discovered with the radial velocity (RV) technique. No transit signatures have been observed for that planet on the original light curves, hence its orbital inclination remains unknown and only a lower constraint of $0.40 \pm 0.05$ $M_{\rm{Jup}}$ can be placed on its mass.

The planets are close to a 1:4 orbital period commensurability that boosts their mutual gravitational interactions. Transit times for WASP-148~b were found to deviate from a linear ephemeris, which is representative for a Keplerian approach, by more than 15 minutes (H\'ebrard \etal 2020). Those transit time variations (TTVs) may place tighter constraints on the orbital parameters -- the key issues in studies of a dynamical state of the system and in tracking its orbital evolution. 

The origin of massive planets on tight orbits, so called hot (with orbital periods $P_{\rm{orb}}< 10$ days) or warm ($10<P_{\rm{orb}}< 200$ days) Jupiters, still remains enigmatic. Those planets are thought to be formed beyond a water frost line and then to migrate inwards due to tidal interactions with a protoplanetary disk in early stages of system evolution (Lin \etal 1996), due to tidal damping of a high orbital eccentricity excited by a planetary or stellar companion (Rasio \& Ford 1996, Eggleton \& Kiseleva-Eggleton 2001), or due to external perturbations induced in a spatially clustered stellar formation environment (Winter \etal 2020). An alternative formation channel that accepts in-situ formation via a core-accretion process is also considered (Batygin \etal 2016).

In a result of high-eccentricity migration, hot giant planets are expected to be devoid of planetary companions with orbital periods shorter than $\approx 100$ days. This formation pathway is supported by observations showing that hot Jupiters are unlikely to exist in compact systems (Wright \etal 2009, Latham \etal 2011, Steffen \etal 2012). At best, they are accompanied by massive planets on wide and eccentric orbits (\eg Bonomo \etal 2017). This points to the high-eccentricity migration as a dominant formation channel of hot Jupiters. There is, however, a handful of exceptions from this picture: WASP-47~b with a Neptune-sized outer planet and a super-Earth inner companion (Becker \etal 2015) and accompanied by a distant massive planet (Neveu-VanMalle \etal 2016), Kepler-730~b with an interior Earth-sized planet (Ca{\~n}as \etal 2019), and TOI-1130~b with a Neptune-sized inner companion (Huang \etal 2020). 

Warm Jupiters, in contrary, are found in compact multi-planetary systems more frequently (Steffen \etal 2012, Huang \etal 2016). This observation suggests that those planets were formed in situ or migrated nonviolently in protoplanetary disc. 

The WASP-148 system with its inner planet on a $\approx 9$ day orbit is on a border line between systems with hot and warm Jupiters. It could be a short-period tail of warm Jupiters rather than a long-period tail of hot gas giants (Huang \etal 2016). Our better understanding of WASP-148's architecture would help answer the question on system's origin and dynamical evolution. In this paper, we construct and explore a self-consistent orbital model of the system in a Newtonian approach. For this, we combined transit timing and radial velocity data, which are available in the literature, with new WASP-148~b's transit times that we extracted from space-borne and ground-based follow-up photometric observations.

\section{Observations and data reduction}

\subsection{TESS photometry}

WASP-148 was observed by the Transiting Exoplanet Survey Satellite (TESS, Ricker \etal 2014) between 2020 April 16 and 2020 July 04 (Sectors 24--26) in a low cadence mode with exposure times of 30 minutes. Post-stamps of a $15 \times 15$ pixel width ($5.25' \times 5.25'$), centred on WASP-148, were extracted from full-frame images with the TESSCut\footnote{https://mast.stsci.edu/tesscut/} online tool (Brasseur \etal 2019). Standard procedures available in the Lightkurve v1.9 package (Lightkurve Collaboration 2018) were employed to produce a photometric time series. Visual inspection of a field around WASP-148 and querying the GAIA DR2 catalog (Gaia Collaboration \etal 2018) revealed no nearby bright stars that could be blended due to the limited angular resolution of TESS cameras. An aperture radius was 2 or 3 pixels with additional manual optimisation of the aperture mask in order to account for TESS inter-sector changes of a stellar profile and orientation of a field of view. An algorithm with standard-deviation thresholding was applied to map the sky background. In-transit and in-occultation data points were masked out taking an ephemeris from H\'ebrard \etal (2020), which is also used for transit numbering throughout this study. Then the Savitzky-Golay filter with a window of 12 hours was applied to remove any trends caused by instrumental effects or stellar variation. The final TESS photometric time series is shown in Fig.~1.

\begin{figure}[thb]
\begin{center}
\includegraphics[width=1.0\textwidth]{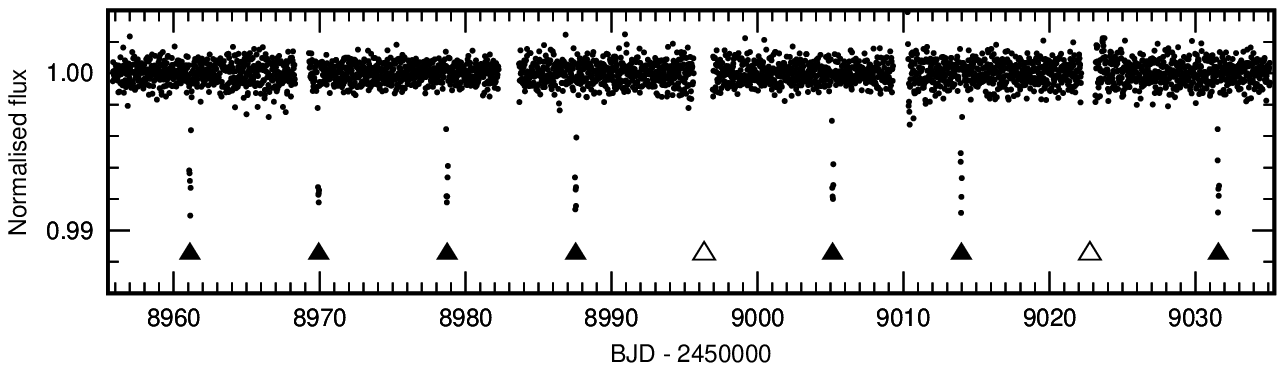}
\end{center}
\FigCap{Normalised TESS light curve for WASP-148 from Sectors 24--26 after de-trending. The observed transits cover epochs 114--122 (counted from the initial transit given by H\'ebrard \etal 2020) and are indicated by filled triangles. Open triangles mark missing transits at epochs 118 and 121 that fell in technical interruptions in observations.}
\end{figure}

Seven complete transits between epochs 114 and 122 were covered with observations. Two transits at epochs 118 and 121 were missed because they happened when TESS's science operation was interrupted. Individual transit light curves around the expected transit mid-points were extracted for further analysis with time margins equal to $\pm2.5$ times a transit duration. Their photometric noise rates (pnr, Fulton \etal 2011) were found to be between 3.0 and 4.0 ppth of the normalised flux per minute of observation.

\subsection{Ground-based photometry}

Two transits of WASP-148~b, including one partial, were observed with the 1.5 m Ritchey-Chr\'etien Telescope at the Sierra Nevada Observatory (OSN, Spain). The instrument was equipped with a Roper Scientific VersArray 2048B CCD camera. The field of view was $7.92' \times 7.92'$. To increase the signal-to-noise ratio for transit timing purposes, no filter was used and the light curves were acquired in white light. 

The first run was executed on 9 July 2020, starting just after nautical twilight to observe the second part of the first transit succeeding the TESS observations. To reduce CCD readout time, a $2 \times 2$ pixel binning was applied. The second run was executed on 22 August 2020 without binning. A longer observing window allowed a complete transit to be recorded. In both runs, the sky was clear but conditions were not perfectly photometric due to fluctuations in atmospheric transparency induced by thin clouds. The instrument was auto-guided and mildly defocused to keep the target and comparison stars below a saturation level. The further details on the observing runs are given in Table~1.

\MakeTable{l l c c c c c}{12.5cm}{Details on the OSN light curves. Date UT is given for the middle of the transit, epoch is the transit number from the initial ephemeris given in H\'ebrard \etal (2020), $X$ shows the airmass change during transit observations, $N_{\rm{obs}}$ is the number of useful scientific exposures, $t_{\rm{exp}}$ is exposure time, $\Gamma$ is the median number of exposures per minute, pnr is explained in the text.}
{\hline
Date UT (Epoch)  & UT start-end  &  $X$       & $N_{\rm{obs}}$ & $t_{\rm{exp}}$ [s] & $\Gamma$ & pnr [ppth]\\
\hline
 2020 Jul 9 (123)  & 20:43--23:50 & $1.04 \rightarrow 1.01 \rightarrow1.08$ & 502 & 20 & 2.69  &  0.85\\
 2020 Aug 22 (128)  & 19:49--00:02 & $1.02 \rightarrow 1.74$ & 594 & 20 & 2.35  &  1.18\\
\hline
}

The science frames were reduced with AstroImageJ software (Collins \etal 2017) following a standard procedure which included de-biasing, flat fielding, and transforming timestamps into barycentric Julian dates in barycentric dynamical time $(\rm{BJD_{TDB}})$. Fluxes were obtained with the differential aperture photometry method with an aperture size and selection of comparison stars optimised to minimise the photometric data scatter. The photometric time series were de-trended against the airmass, time, and seeing along with a trial transit model. Then the fluxes were normalised to unity out of the transits. The light curves in machine-readable form are available via CDS.

\section{Data analysis and results}

\subsection{Modelling of transit light curves}

To refine parameters of the WASP-148~b's transits, the TESS and OSN light curves were modelled simultaneously using the Transit Analysis Package (TAP, Gazak \etal 2012). The software uses the Mandel \& Agol (2002) approximation to model a transit geometry and observed flux, and employs the Markov Chain Monte Carlo (MCMC) technique driven by the Metropolis-Hastings algorithm and a Gibbs sampler to find the best-fitting solution and parameters' uncertainties. We took a standard approach with 10 random walk chains, each $10^6$ steps long with 10\% burn-in phase. The median values of the posteriori parameter distributions were taken as the best-fitting values. Their uncertainties were determined from 15.9 and 84.1 percentiles of the marginalised posteriori probability distributions.

A transit light curve model was characterised by an orbital inclination $i_{\rm{b}}$, a semi-major axis scaled in stellar radii $a_{\rm{b}}/R_{\star}$, a ratio of planet to star radii $R_{\rm{b}}/R_{\star}$, limb darkening (LD) coefficients, a mid-transit time $T_{\rm{mid}}$, and possible trends in the time domain. The first three of those parameters were linked together for all light curves. In a test run, $R_{\rm{p}}/R_{\star}$ in TESS and OSN passbands were allowed to be determined independently to account for third light contamination\footnote{We note, however, that the host star appears to be well separated in TESS images.} or LD inaccuracy. Both values however were found to be consistent within a 1$\sigma$ range that justifies the final approach. The mid-transit times were determined for each transit light curve separately. They are given in Table~2.

\MakeTable{c c c}{12.5cm}{Mid-transit times taken from literature and determined for the new transit light curves.}
{\hline
Epoch  & $T_{\rm{mid}}$ $[\rm{BJD_{TDB}}-2450000]^{*}$  &  Data source\\
\hline
$-377$ & $4638.453766 \pm 0.026000$ & H\'ebrard \etal (2020) \\
$-296$ & $5351.525766 \pm 0.017000$ & H\'ebrard \etal (2020) \\
$-255$ & $5712.519266 \pm 0.006100$ & H\'ebrard \etal (2020) \\
$-122$ & $6883.444266 \pm 0.001400$ & H\'ebrard \etal (2020) \\
$-87$ & $7191.557466 \pm 0.000820$ & H\'ebrard \etal (2020) \\
$-46$ & $7552.505266 \pm 0.001200$ & H\'ebrard \etal (2020) \\
$-11$ & $7860.647466 \pm 0.001300$ & H\'ebrard \etal (2020) \\
  0 & $7957.481536 \pm 0.000300$ & H\'ebrard \etal (2020) \\
114 & $8961.122300^{+0.002922}_{-0.002914}$ & TESS \\
115 & $8969.927527^{+0.002350}_{-0.002184}$ & TESS \\
116 & $8978.735938^{+0.002190}_{-0.002248}$ & TESS \\
117 & $8987.540311^{+0.003212}_{-0.002834}$ & TESS \\
119 & $9005.153892^{+0.002386}_{-0.002353}$ & TESS \\
120 & $9013.955938^{+0.003285}_{-0.003242}$ & TESS \\
122 & $9031.570109^{+0.002784}_{-0.002746}$ & TESS \\
123 & $9040.372302^{+0.001551}_{-0.001727}$ & OSN \\
128 & $9084.397443^{+0.000848}_{-0.000849}$ & OSN \\
\hline
\multicolumn{3}{l}{$^{*}$ higher numerical precision left intentionally.} \\
}

The LD effect was approximated with the quadratic law (Kopal 1950) using the linear $u_{\rm 1}$ and quadratic $u_{\rm 2}$ coefficients. Their values were taken from the tables of Claret \& Bloemen (2011) as averages of $R$, $I$, and Sloan Digital Sky Survey $z'$ values for the TESS passband and averages of $B$, $V$, $R$, and $I$ values for the OSN clear passband. The stellar parameters, such as the effective temperature, gravity acceleration, and metallicity were taken from H\'ebrard \etal (2020). In the fitting procedure, the LD coefficients were allowed to vary around their theoretical values under a Gaussian penalty with a conservative value of 0.1.

The trends in the time domain were determined for each photometric time series separately. They were approximated with a second-order polynomial. Although the polynomial's coefficients were found to be consistent with zero well within a 1$\sigma$ range for almost all light curves, their uncertainties contributed to a total error budget of the fit in order to account for de-trending applied during data reduction. 

The values of the orbital period $P_{\rm{b}}$, eccentricity $e_{\rm{b}}$, and argument of periastron $\omega_{\rm{b}}$ were taken from a Newtonian model of the WASP-148 system (Sect.\ 3.2). To account for their uncertainties in the total error budget of the fit they were allowed to vary under Gaussian penalties determined by their 1$\sigma$ errors.

The redetermined transit parameters were used to calculate a transit depth: 
\begin{equation}
     \delta = \left( R_{\rm{b}}/R_{\star} \right)^2 \, ,
\end{equation}
a transit impact parameter normalised to the stellar radius:
\begin{equation}
     b = \frac{a_{\rm{b}}}{R_{\star}} \cos(i_{\rm{b}}) \left( \frac{1-e_{\rm{b}}^2}{1+e_{\rm{b}} \sin \omega_{\rm{b}}} \right) \, ,
\end{equation}
a total transit duration (from the first to the fourth contact):
\begin{equation}
     T_{14} = \frac{P_{\rm{b}}}{\pi} \left( \frac{a_{\rm{b}}}{R_{\star}} \right)^{-1} \sqrt{(1+R_{\rm{b}}/R_{\star})^2 - b^2} \, ,
\end{equation}
and a mean stellar density using the Kepler's third law under the assumption that the planet's mass is negligible compared to the mass of the host star: 
\begin{equation}
     \rho_{\star} = \frac{3\pi}{G P_{\rm{b}}^2} \left(\frac{a_{\rm{b}}}{R_{\star}}\right)^3\, .
\end{equation}
The results are listed in Table~3. The individual light curves with the best-fitting model and the residuals are plotted in Fig.~2.

\MakeTable{ l c c }{12.5cm}{Parameters for the WASP-148 system obtained from modeling of WASP-148~b's transit light curves. The results from H\'ebrard \etal (2020) are given for comparison purposes.}
{\hline
 Parameter & This paper & H\'ebrard \etal (2020) \\
\hline
Orbital inclination, $i_{\rm{b}}$ $[^{\circ}]$ & $88.53^{+0.90}_{-0.82}$      & $89.80\pm0.27$ \\
Scaled semi-major axis, $a_{\rm{b}}/R_{\star}$ & $19.8^{+2.3}_{-2.6}$         & $19.8 \pm 1.5$ \\
Radii ratio, $R_{\rm{b}}/R_{\star}$            & $0.0837^{+0.0034}_{-0.0026}$ & $0.0807 \pm 0.0007$ \\
Transit depth, $\delta$ [ppth]                 & $7.01^{+0.57}_{-0.44}$       & $6.51 \pm 0.11$ \\
Impact parameter, $b$ $[R_{\star}]$            & $0.482^{+0.096}_{-0.099}$    & $0.046 \pm 0.066$ \\
Transit duration, $T_{14}$ $[h]$               & $3.30^{+0.42}_{-0.47}$       & $3.016 \pm 0.019$ \\
Stellar density, $\rho_{\star}$ $[\rho_{\odot}]$ & $1.34^{+0.47}_{-0.53}$     & $1.34 \pm 0.34$ \\
\hline
}

\begin{figure}[thb]
\begin{center}
\includegraphics[width=1.0\textwidth]{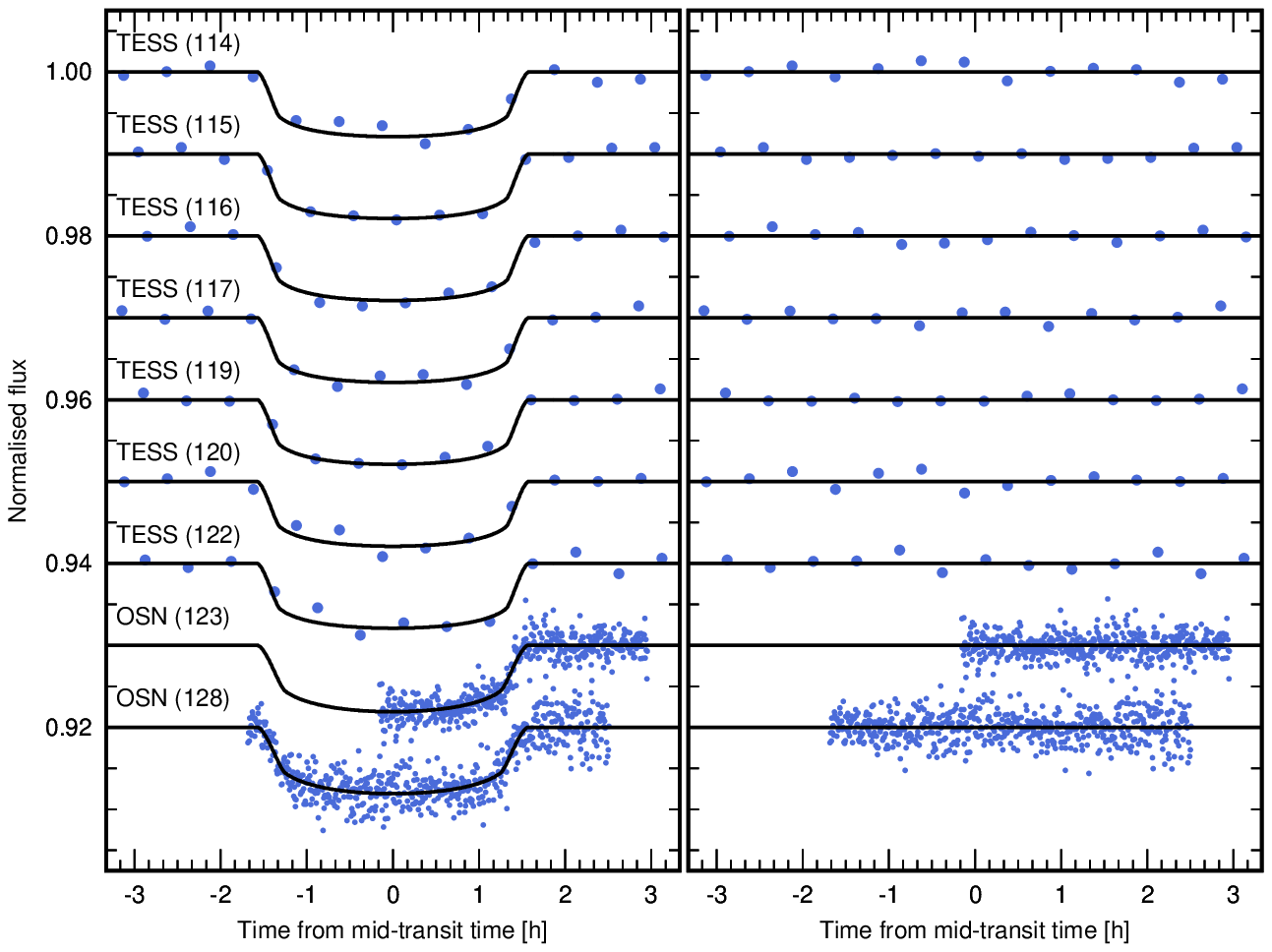}
\end{center}
\FigCap{New transit light curves from TESS and OSN with the best-fitting model are shown in left panel. Epochs (transit numbers) that identify individual transits are given in parentheses. The residuals are plotted in right panel.}
\end{figure}

\subsection{Newtonian model of the system}

The Systemic software (version~2.182, Meschiari \etal 2009) was used to find the best-fitting Newtonian model of the system. We used 116 RV measurements from H\'ebrard \etal (2020) which were acquired between April 2014 and June 2018 with the SOPHIE spectrograph at the 1.93 m telescope of the Observatoire Haute-Provence (France). The transit-times dataset included 8 mid-transit times from H\'ebrard \etal (2020) spanning 9 years from June 2008 to July 2017 and 9 new mid-transit times from Sect.\ 3.1. All time stamps were converted into $\rm{BJD_{TDB}}$. 

We started with the construction of a coplanar Keplerian model with 2 planets reproducing a solution presented by H\'ebrard \etal (2020). The Levenberg-Marquardt (LM) method was used to find the best-fitting solution that was then taken as a starting point in search for the Newtonian model. The Runge-Kutta (RK89) algorithm was employed to integrate equations of motion with an accuracy requirement of $10^{-16}$. The assumption of coplanarity was upheld. The best-fitting solution was found with the differential evolution algorithm, which is implemented in Systemic, followed by LM optimisation. Parameters' $1\sigma$ uncertainties were estimated with the bootstrap method with $10^4$ trials as median absolute deviations.

The results are given in Table~4. In addition to the values of the orbital periods $P$, eccentricities $e$, and arguments of periastron $\omega$, there are the true anomalies $\nu$ and RV amplitudes $K$ for both planets, barycentric systemic RV $\gamma$, and stellar jitter. The orbital parameters are given for epoch $\rm{BJD_{TDB}}$ = 2459048.747 (July 2020) that is the average time of the TESS and OSN transit observations weighted with the timing uncertainties. This approach provides the most representative values of $P_{\rm{b}}$, $e_{\rm{b}}$, and $\omega_{\rm{b}}$ that were used in transit light curve modeling (Sect.\ 3.1). We note that changes in these parameters due to dynamical evolution of the system can be neglected in the time window spanning the TESS and OSN observations because they are much smaller than the parameters' uncertainties.  

\MakeTable{ l c c c c }{12.5cm}{Parameters for the WASP-148 system in the Newtonian approach, given for  epoch $\rm{BJD_{TDB}}$ = 2459048.747. The Keplerian model from H\'ebrard \etal (2020) is given for comparison purposes.}
{\hline
 Parameter & \multicolumn{2}{c}{Newtonian model} & \multicolumn{2}{c}{Keplerian model (H\'ebrard \etal 2020)} \\
           & planet b & planet c & planet b & planet c \\
\hline
$P$ $[d]$ & $8.80464 \pm 0.00054$ & $34.541 \pm 0.013$ & $8.803810 \pm 0.000043$ & $34.516 \pm 0.029$\\
$e$ & $0.080 \pm 0.014$ & $0.29 \pm 0.03$ & $0.220 \pm 0.063$ & $0.359 \pm 0.086$\\
$\omega$ $[^{\circ}]$ & $37 \pm 10$ & $7 \pm 13$ & $59 \pm 20$ & $14 \pm 17$\\
$\nu$ $[^{\circ}]$ & $33 \pm 10$ & $152 \pm 6$ & $-$ & $-$\\
$K$ $[\rm{m \, s^{-1}}]$ & $28.6 \pm 1.9$ & $25.5 \pm 2.1$ & $28.7 \pm 2.0$ & $25.9 \pm 2.9$\\
$\gamma$ $[\rm{m \, s^{-1}}]$ & \multicolumn{2}{c}{$-5617.5 \pm 1.3$} & \multicolumn{2}{c}{$-5619 \pm 5$}\\
jitter $[\rm{m \, s^{-1}}]$ & \multicolumn{2}{c}{$10.6$} & \multicolumn{2}{c}{$11.1 \pm 1.4$}\\
\hline
}

Compared with the Keplerian model of H\'ebrard \etal (2020), our Newtonian model puts tighter constraints on the orbital period of  planet c $P_{\rm{c}}$ thanks to accounting for gravitational interaction with planet b. On the other hand, $P_{\rm{b}}$ was found to be constrained one order of magnitude weaker. This is a consequence of the influence of the uncertainties of other orbital parameters  on the error budget of $P_{\rm{b}}$, making our error estimates more reliable.

To investigate a role of the new timing data in better understanding of the system's parameters, we searched for a trial solution using only data from H\'ebrard \etal (2020). This exercise showed that the uncertainties of the key parameters, i.e.\ the orbital periods and eccentricities, were reduced by a factor 3--4 thanks to our new observations. 

Transit timing variations for planet b are shown in Fig.~3. Their range is $\approx$40 minutes with a dominant period of $\approx$450 days. Our transit times cover an ascending branch and maximum of that signal.

\begin{figure}[thb]
\begin{center}
\includegraphics[width=1.0\textwidth]{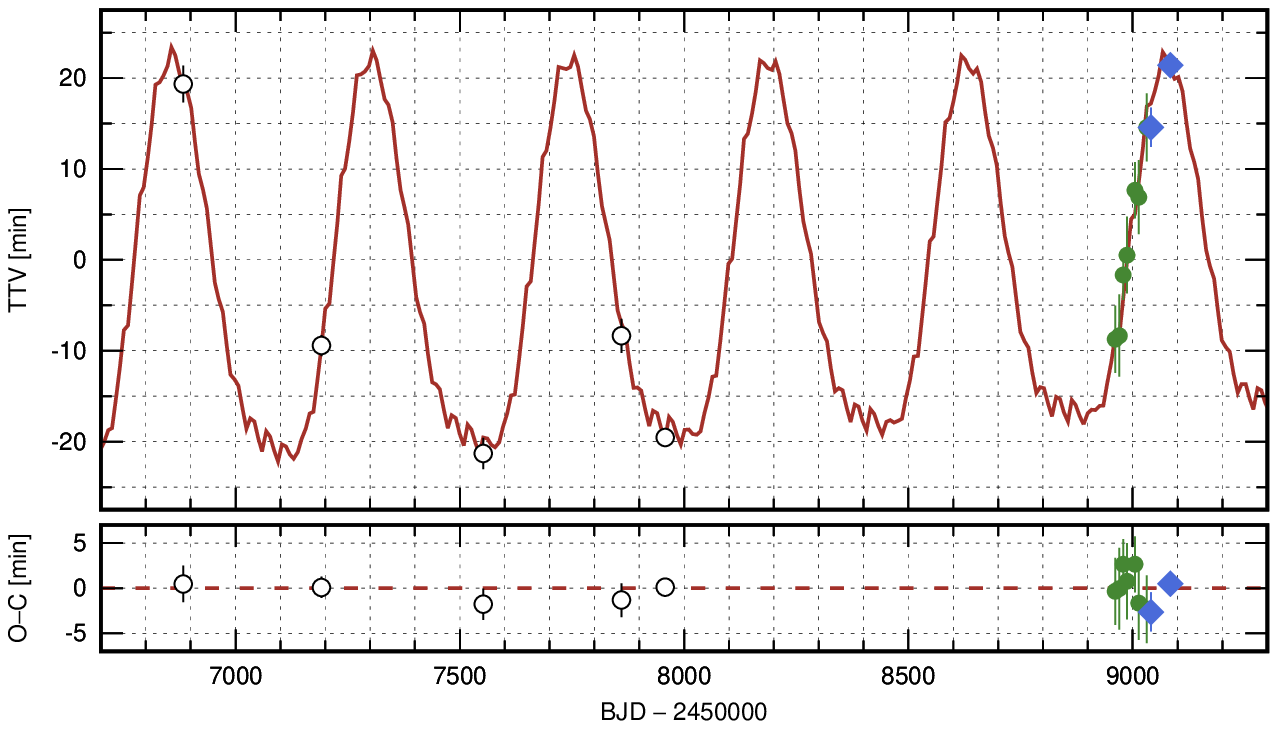}
\end{center}
\FigCap{Upper panel: transit timing variations for WASP-148~b. The new determinations are denoted by dots: green for TESS, blue for OSN. Data from literature are marked with open symbols. The continuous line shows the TTVs predicted by our Newtonian model. Lower panel: the final transit timing residuals. Three low-precision data points, which were extracted from SuperWASP photometry by H\'ebrard \etal (2020), are out of the plot range.}
\end{figure}

The model data were utilised to provide an approximated transit ephemeris in a form:
\begin{equation}
     T_{{\rm mid}} \, [{\rm BJD_{TDB}}] = 2457957.493 + 8.80381 \cdot E + 0.014\cdot \sin \left[ \frac{2\pi}{50.16} (E-14.62) \right],
\end{equation}
where $E$ is the transit number. This ephemeris can be useful in planning follow-up observations of WASP-148~b. Its linear component represents the Keplerian solution and is accurate up to $\approx$ 20 minutes. The periodic component is a first-order approximation of the Newtonian model and assures an  accuracy up to 5 minutes. 

To verify dynamical stability of our model, we conducted a numerical experiment in which we tracked the evolution of the system parameters over $10^6$ years. We employed the RK89 integrator with a step of $0.01 P_{\rm{b}}$. The system was found to be stable, and exemplary variations of eccentricities over a period of 5000 years are plotted in the upper panel in Fig.~4. They oscillate between 0.06 and 0.33 for planet b and between 0.20 and 0.30 for planet c in an anti-correlated manner. A period of those variations is $\approx 1000$ years, i.e. is longer than 881 years reported by H\'ebrard \etal (2020). The ratio of orbital periods $P_{\rm{c}}/P_{\rm{b}}$ is close to the 4:1 commensurability. To check whether the planets are trapped in a mean motion resonance we tracked the evolution of the difference between arguments of periastron, defined as $\Delta\omega=\omega_{\rm{b}}-\omega_{\rm{c}}$. As it is illustrated in the bottom panel in Fig.~5, $\Delta\omega$ shows librations around $0^{\circ}$ with an amplitude of $44^{\circ}$ which is slightly less than $45^{\circ}$ reported by H\'ebrard \etal (2020). Those oscillations point to an apsidal alignment of both orbits. Such conditions are generated by linear secular coupling and prevent both planets from close encounters that could lead to destabilisation of the system. However, none of the eccentricity-type resonant angles, defined as a linear combination of
mean longitudes $\lambda$ and arguments of periastron,
\begin{align}
 \theta_{1} & = \lambda_c - 4\lambda_{\rm{b}}+\omega_{\rm{b}}+2\omega_{\rm{c}} \\
 \theta_{2} & = \lambda_c - 4\lambda_{\rm{b}}+2\omega_{\rm{b}}+\omega_{\rm{c}} \\
 \theta_{3} & = \lambda_c - 4\lambda_{\rm{b}}+3\omega_{\rm{b}} \\
 \theta_{4} & = \lambda_c - 4\lambda_{\rm{b}}+3\omega_{\rm{c}}
\end{align}
was found to librate, showing that the planets are out of the dynamical resonance 4:1.

\begin{figure}[thb]
\begin{center}
\includegraphics[width=1.0\textwidth]{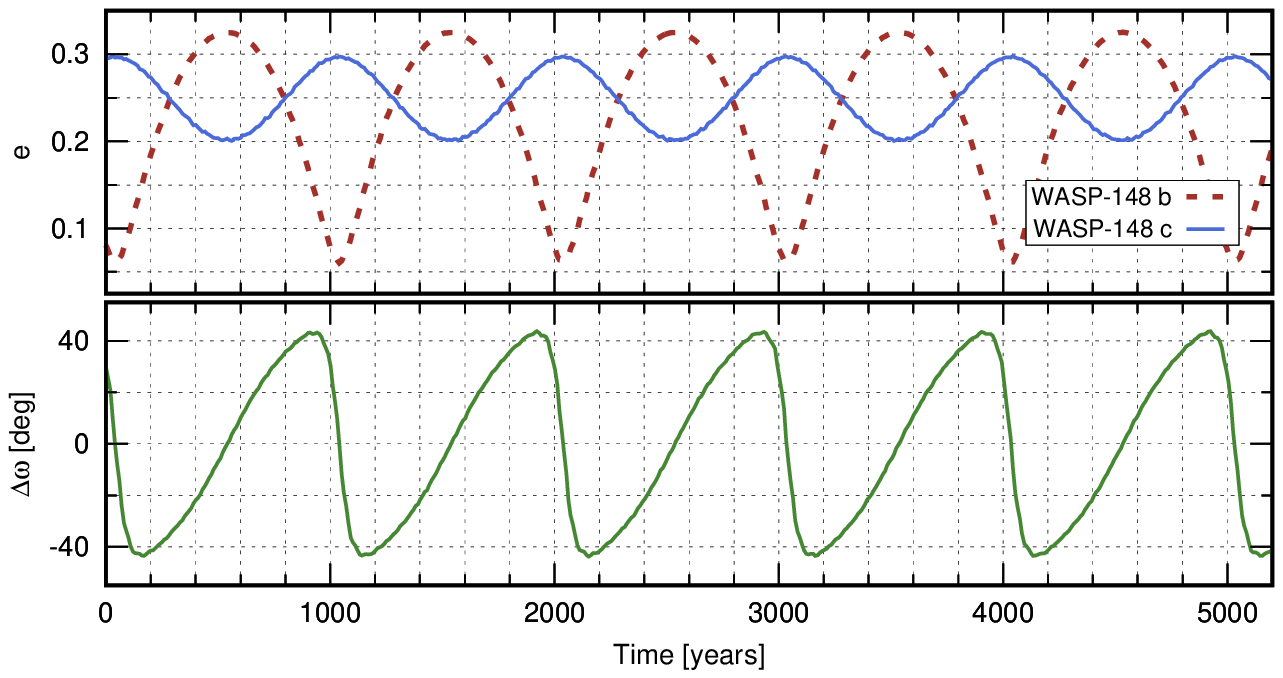}
\end{center}
\FigCap{Upper panel: evolution of the orbital eccentricities (upper panel) and the difference between the arguments of periastron (lower panel) for the WASP-148 planets as predicted by our Newtonian model over the next 5000 years.}
\end{figure}

Switching from the Keplerian to Newtonian model brings slight deterioration of the goodness of the fit in the RV domain. The Keplerian model yields $\chi^2_{\rm{RV}} = 244.3$ and the stellar jitter of 10.3 $\rm{m \, s^{-1}}$, while for the Newtonian one these parameters are 253.0 and 10.6 $\rm{m \, s^{-1}}$, respectively. H\'ebrard \etal (2020) noticed that in the RV residuals of the two-planet model there is a 150-day periodic signal which remains just below an acceptance level of false alarm probability (FAP) equal to 0.1\%. It was identified as a possible signature of a third outer planet. While reproducing this result, we noticed that the significance of the 150 day signal is much lower. For the RV residuals, a periodogram for a range of periods between 0.5 and $10^4$ days was generated using the analysis of variance algorithm (AoV, Schwarzenberg-Czerny 1996) with 1 harmonic. The RV uncertainties were used to calculate weights. Levels of FAP were estimated with the bootstrap method based on $10^4$ trials. The 150 day signal was found to be the strongest peak with FAP of 22.1\% and 14.5\% for Keplerian and Newtonian models, respectively. 

Since the 150 day signal seems to be strengthened in the Newtonian approach, we tested a trial model with this third planet injected. When compared to the two-planet model, there is improvement in $\chi^2_{\rm{RV}} = 207.4$ leaving a jitter value of 8.4 $\rm{m \, s^{-1}}$. The same observation applies to transit timing: the value of $\chi^2_{\rm{TT}}$ drops from 13.6 down to 8.4. This noticeable improvement is due to a long-term trend in WASP-148~b's TTVs that is induced by the third planet. However, the Bayesian information criterion (BIC), defined as ${\rm{BIC}} = {\chi}^2 + k \ln N$, where $k$ is the number of fit parameters and $N$ is the number of data points, strongly disfavours the three-planet model with $\Delta \rm{BIC} = {\rm{BIC}}_{\rm{[3 \, planets]}} - {\rm{BIC}}_{\rm{[2 \, planets]}} \approx 96$ due to a greater number of fitted parameters. 

In a numerical experiment, we allowed $i_{\rm{c}}$ and the relative longitude of the nodes $\Delta \Omega = \Omega_{\rm{c}} - \Omega_{\rm{b}}$ to be the free parameters of a trial model. For simplicity, the value of $\Omega_{\rm{b}}$ was fixed at $0^{\circ}$, hence $\Delta \Omega = \Omega_{\rm{c}}$. The best-fitting model, stable on $10^6$ yr time scales, was found for $\Delta \Omega \approx -17^{\circ}$ and $i_{\rm{c}} = 47^{\circ}$ (and by symmetry $\Delta \Omega \approx +17^{\circ}$ and $i_{\rm{c}} = 133^{\circ}$) with a marginal improvement of $\chi^2$. Qualitatively similar results were obtained by H\'ebrard \etal (2020). We note, however, BIC disfavours this trial model with $\Delta \rm{BIC} = {\rm{BIC}}_{\rm{[non-coplanar]}} - {\rm{BIC}}_{\rm{[coplanar]}} \approx 97$. Thus we recognise the present datasets as being insufficient to study the orbital non-coplanarity in the system.

\subsection{Search for transits of WASP-148~c}

The coplanar Newtonian model (Sect.~3.2) predicts that the transit impact factor of WASP-148~c is $1.1 \pm 0.3$, leaving within a 1$\sigma$ range the possibility that transits of this planet could be observable. The transit ephemeris was found to be accurate to 0.7 d ($\approx 16$ hours) and predicts two time windows around BJD 2458971.9 and 2459006.4 that are covered with TESS photometry. As it is shown in Fig.~5, there is no sign of a transit signature that could be identified by a visual inspection. According to the Extrasolar Planets Encyclopaedia\footnote{http://exoplanet.eu} (Schneider \etal 2011), planets with masses similar to the mass of WASP-148~c, i.e. $\approx 0.4$ $M_{\rm{Jup}}$, have their radii between 0.4 and 1.6 $R_{\rm{Jup}}$. If WASP-148~c had the size from that range, it would produce transits 2.1--33 ppth deep. As it is illustrated in Fig.~5, even the shallowest transits could be identified by a visual inspection. 

\begin{figure}[thb]
\begin{center}
\includegraphics[width=1.0\textwidth]{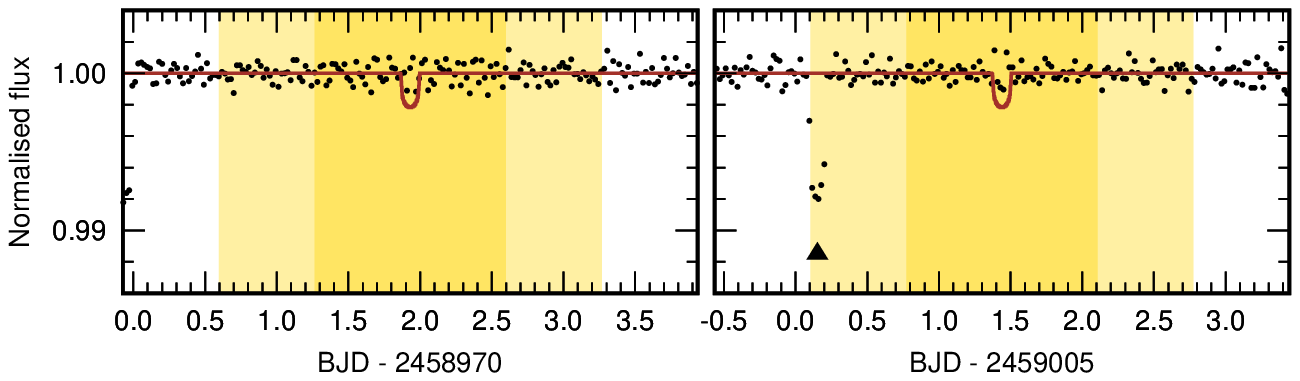}
\end{center}
\FigCap{TESS photometric time series around two time windows in which transits of WASP-148~c's are predicted by the coplanar Newtonian model. The shaded areas show 1- and 2-$\sigma$ uncertainties of the transit ephemeris. A model of a hypothetical transit for a 0.4 $R_{\rm{Jup}}$ planet with the orbital parameters of WASP-148~c and the transit impact factor of 0.8 is plotted with a continuous line.  A black triangle marks a transit of WASP-148~b.}
\end{figure}

\subsection{Search for additional transits}

The TESS photometric time series was searched for other transiting planets. The transits and occultations of WASP-148~b were masked out with a margin of $3 \sigma$ of the transit duration. Box-like periodic signals were searched with the AoV method optimised for transit signals (AoVtr, Schwarzenberg-Czerny \& Beaulieu 2006). The algorithm folds with a trial period and bins the light curve to test a negative-pulse model with a minimum associated to a transit event. Following Maciejewski (2020), we used 100 bins providing an optimised resolution for the trial periods from a range between 0.5 and 100 days. A periodogram resolution in frequency was $2.5 \times 10^{-4}$ day$^{-1}$. The bootstrap method was employed to empirically determine FAP levels. 

The periodogram is displayed in the upper panel of Fig.~6. As it can be seen, there is no statistically significant signal. The highest peak, placed at $\approx 0.72$ d, was a subject of a careful visual inspection which definitely rejected it as a transit-shape signature. There is also no signal around $P_{\rm{c}}$ that strengthens the results of the performed visiual inspection of the light curve.

Using the procedure adopted from Maciejewski (2020), an upper constraint on a depth of transits that remain below a detection threshold was determined as a function of orbital periods of hypothetical planets. Artificial transit signals with depths starting from 0.01 ppth were injected into the TESS photometric time series, and then a trial AoVtr periodogram was computed. The depth of the synthetic transits was iteratively increased with a step of 0.01 ppth until the power peak associated to a tested period reached the FAP level of 0.1\% which assures an unquestionable transit detection. 

The results are shown in the lower panel of Fig.~6. Transits deeper than $\approx 0.6$ ppth would be detected for periods shorter than $P_{\rm{b}}$. This detection threshold translates into planetary radii of $\approx 2.4$ $R_{\oplus}$, allowing us to probe down to the regime of mini-Neptunes (Buchhave \etal 2014). The detection threshold gradually increases up to a few ppth for the longer periods. For periods close to $P_{\rm{c}}$, the threshold value is $\approx2.2$ ppth which is similar to the depth of the shallowest transits expected for WASP-148~c from comparative planetology.

\begin{figure}[thb]
\begin{center}
\includegraphics[width=0.714\textwidth]{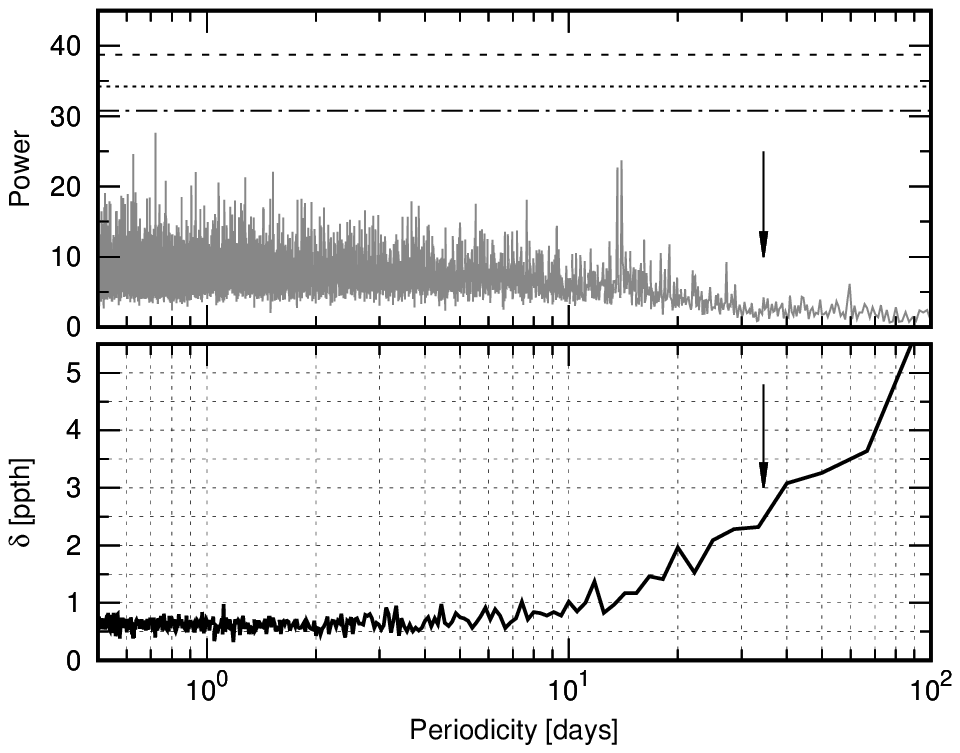}
\end{center}
\FigCap{Upper panel: AoVtr periodogram for TESS photometric time series with signatures of WASP-148~b masked out. The dashed lines place the empirical FAP levels of 5\%, 1\%, and 0.1\% (from the bottom up). The orbital period of WASP-148~c is marked with an arrow. Lower panel: upper constraints on depths of hypothetical transits that remain under the detection threshold.}
\end{figure}

\section{Conclusions}

The new transit times for WASP-148~b allowed us to refine parameters of the Newtonian model of the WASP-148 system. The uncertainties of the orbital periods and eccentricities were reduced by a factor of 3--4. The planet b was found to have the eccentricity lower than the value reported by H\'ebrard \etal (2020). Its value for epoch $\rm{BJD} = 2459048.747$ (July 2020) is $\approx 0.08$ and is predicted to decrease down to $\approx 0.06$ around 2060. Then it will increase up to $\approx 0.33$ around 2560.

Planet b exhibits the pronounced TTV with the amplitude of $\approx 20$ minutes which can be  followed-up with ground-based 1-m class telescopes. Since periodicity of this signal is $\approx 450$ days, new observations in 2021 are expected to cover its phase around minimum, i.e.\ with transits recorded up to $\approx 20$ min earlier than a linear ephemeris predicts.

Planet c does not transit and this finding is in line with coplanarity of the system. Although current observations do not allow the mutual inclinations to be determined, the stability studies show that it must be lower than $\approx 35^{\circ}$ (H\'ebrard \etal 2020). 

The RV measurements point to the possible existence of the third planet on the 150-day orbit. Although introducing it into the Newtonian model improves the goodness of the fit in both the Doppler and TTV domains, the statistical significance of the signal remains below the acceptance level. Additional RV observations may help to clarify this issue.

No additional transiting planets have been found in the system. The TESS photometry allowed us to probe down to $\approx 2.4$ $R_{\oplus}$ for interior planets. We note that the regime of rocky planets still remains unexplored, and planets similar in size to WASP-47~e ($\approx 1.8$ $R_{\oplus}$, Becker \etal 2015) evade being discovered. Such planets could be detected with the CHEOPS space telescope (Broeg \etal 2013). That makes the WASP-148 system an interesting target in further observations.


\Acknow{GM acknowledges the financial support from the National Science Centre, Poland through grant no. 2016/23/B/ST9/00579. MF acknowledges financial support from grant AYA2016-79425-C3-3-P of the Spanish Ministry of Economy and Competitiveness (MINECO), co-funded with EU FEDER funds, and grant PID2019-109522GB-C5X/AEI/10.13039/501100011033 of the Spanish Ministry of Science and Innovation (MICINN). MF, AS, and AGS acknowledge financial support from the State Agency for
Research of the Spanish MCIU through the \textit{Center of Excellence Severo
Ochoa} award to the Instituto de Astrofisica de Andalucia (SEV-2017-0709). This paper includes data collected with the TESS mission, obtained from the MAST data archive at the Space Telescope Science Institute (STScI). Funding for the TESS mission is provided by the NASA Explorer Program. STScI is operated by the Association of Universities for Research in Astronomy, Inc., under NASA contract NAS 5-26555. This research made use of Lightkurve, a Python package for Kepler and TESS data analysis (Lightkurve Collaboration, 2018). This research has made use of the SIMBAD database and the VizieR catalogue access tool, operated at CDS, Strasbourg, France, and NASA's Astrophysics Data System Bibliographic Services.}


\end{document}